\documentclass[prd,showpacs,showkeys]{revtex4}
\usepackage{mathtools}
\usepackage{amsmath}
\usepackage{amssymb}
\usepackage{wasysym}

\usepackage{wasysym}
\usepackage{graphics}
\usepackage{eurosym}
\usepackage{amsfonts}
\usepackage{mdframed}
\usepackage{graphicx}
\usepackage[font={footnotesize,it}]{caption}

\begin{document}

\title{The Bekenstein-Hawking Corpuscular Cascading from the Backreacted Black Hole}

\author{Ali \"{O}vg\"{u}n}
\email{ali.ovgun@pucv.cl}

\affiliation{Instituto de F\'{\i}sica, Pontificia Universidad Cat\'olica de
Valpara\'{\i}so, Casilla 4950, Valpara\'{\i}so, Chile}

\affiliation{Physics Department, Eastern Mediterranean University, Famagusta,
Northern Cyprus, Turkey}

\date{\today }
\begin{abstract}
Exciting peculiarities of Planck-scale physics have immediate
effects on the Bekenstein-Hawking radiation emitted from black holes (BHs). In this
paper, using the tunneling formalism, we determine the Bekenstein-Hawking temperature
for the vector particles from a backreacted black hole (BBH)
constructed from a conformal scalar field surrounded by a BTZ (Banados-Teitelboim-Zanelli) BH. Then, under the
effect of the generalized uncertainty principle, we extend our calculations for scalar particles to understand the
effects of quantum gravity. Then, we calculate an evaporation time for the BBH, the total number of Bekenstein-Hawking particles, and the quantum corrections of the number. We observe that remnants of the BH evaporation occur
and that they affect the Bekenstein-Hawking temperature of the BBH as well as the total
number of Bekenstein-Hawking particles. 

\end{abstract}
\maketitle

\section{Introduction}

Many years ago, Bekenstein suggested that in a quantum theory of gravity, the surface area of a black hole (BH) should have a discrete spectrum\cite{bekenstein3}. In 1972, Bekenstein wrote a seminal paper showing that the entropy of a BH is exactly proportional to the size of its event horizon; then, he showed that there is a maximum amount of information that can be stored in a finite region of space, a concept known as the Bekenstein bound \cite{bekenstein1,bekenstein2}. Following Bekenstein's insights, some of the most significant ideas in theoretical physics were born, such as Bekenstein-Hawking radiation \cite{hawkin1,hawkin2,bekenstein1,bekenstein2}, the BH information paradox, the holographic principle, and  the BH firewall paradox. It is a remarkable fact that, according to the seminal works of Bekenstein and then Hawking
\cite{hawkin1,hawkin2,bekenstein1,bekenstein2}, BHs are not entirely black. That surprising
claim was made over forty years ago. Examining the behavior
of quantum fluctuations around the event horizon of a BH, Bekenstein and Hawking
substantiated the theory that BHs emit thermal radiation with
a constant temperature (the so-called Bekenstein-Hawking temperature) that is directly proportional to the surface gravity $\kappa$, which is the gravitational acceleration
experienced at the BH's horizon \cite{hawkin2,bekenstein1,bekenstein2}:

\begin{equation}
T_{H}=\frac{\hslash\kappa}{2\pi}.\label{1}
\end{equation}
When BHs evaporate, their temperatures increase adiabatically
as a function of their remaining mass. Quantum fluctuations create a
virtual particle pair near the BH horizon. While the particle with
negative energy tunnels into the horizon (absorption), the other having positive energy flies off into spatial infinity (emission)
and produces Bekenstein-Hawking radiation (BHR). There are various methods for calculating the level of BHR, and the two most popular tunneling methods are the null geodesic
method and the Hamilton-Jacobi (HJ) method
\cite{krs1,krs2,krs3,krs4,krs5,krs6,sri1,sri2,ahmed,aa,kimet,vanzo,ma0,ma1,ma2,ma3,aa1,aa2,aa3,aa4,aa5,aa6,aa66,aa666,aa6666,aa66666,aa666666,sh21,aa7,aa8,aa9,aa10,aa11,aa12}.
Both approaches use the tunneling method by applying
the WKB (Wentzel-Kramers-Brillouin) approximation and finding the emission and absorption probabilities of
the tunneling particles, which give the tunneling rate $\Gamma$\ as follows \cite{aa13,aa14,aa15,hu1,hu2,hu3,singleton1,singleton2,singleton3,singleton4,Borun,singleton5}:

\begin{equation}
\Gamma=\frac{P_{emission}}{P_{absorption}}=e^{-2ImS}=e^{-\frac{E}{T}},\label{2}
\end{equation}

where $S$ is the action of the classically forbidden trajectory of
the tunneling particle, which has an energy $E$ and a temperature
$T$. 
Here, the conservation of energy plays an important role. First, there is a transition between states with the same total energy; then, it radiates when the mass of the residual hole must go down. Also, there is a way to lower the mass of the BH. This brings us to the idea that BHs are thought to be excited states in quantum gravity, so that one can derive the BH temperature, which is related to the Boltzmann factor, using the imaginary part of the action during the process of the emission of an s-waves from the inside to the outside of the horizon \cite{krs1,krs2,krs3,krs4,krs5,krs6}. For calculations of the tunneling probability, it is simple to use a WKB approximation. It should be noted that one can ignore the higher-order terms (and use only a linear-order expansion), which have a self-interaction effect that results from the conservation of energy \cite{ma0,ma1,ma2,ma3}. One of the methods for finding $S$ is the HJ method.
This method is generally employed by substituting a suitable ansatz,
with consideration of the symmetries of the spacetime, into the relativistic
HJ equation \cite{aa2,aa3,aa11,aa12,aa13,ali,rahman}. The resulting
radial integral always possesses a pole located at the event horizon.
However, using the residue theory, the associated pole can be analytically avoided.
It is argued that BHs emit radiation that is dissimilar to emitted radiation from a thermal objects except that the BH radiation spectrum is semiclassically sparse.  At least in the weak regime, a quantum BH retains a discrete profile that may result from the simplicity of the spectrum. Different complicated spectra can be obtained, and they cause the radiation to look continuous enough in profile to use different quantum theories \cite{bbh0}. Furthermore, from the emission spectrum of the BH radiation, if the grey body factor is ignored, semiclassical treatment shows that BHs radiate and that the resulting emission spectrum is similar to that of the thermal radiation of a blackbody. Page \cite{cas} showed that there is a very long time gap between the emissions; moreover, the same conclusion, that the cascade of BHR is very sparse was also shown. For an observer at the asymptotic infinity, the flux of BHR from a BH is intensely sparse \cite{viser2}, especially for a 3+1 Schwarzschild BH. However, for dimensional cases that are lower/higher than 3+1 dimensions, Bekenstein-Hawking cascades have a continuous character \cite{bekenstein3,bekenstein4,Makela,hod1,hod2}.

The collapsed pure initial state supports the formation of a BH, which was concluded by Hawking, and then the radiation shades into a high-entropy mixed state. This is contrary to the fundamentals of quantum mechanics which can not maintain its pure state, so it becomes mixed up and causes an information paradox \cite{hod1}.
Here, we use a backreacted black hole (BBH), which is a BTZ (Banados-Teitelboim-Zanelli) BH surrounded by a conformal scalar field \cite{btz,bbh0,bbh1,bbh2,bibhas}, to investigate the BHR from tunneling vector particles and quantum gravity-affected scalar particles.
Our motivation for working on 2+1 dimensions is to make the problem much
easier. First, there are no propagating degrees of freedom.
Quantum gravity in 2+1 dimensions is renormalizable and finite. It is also well known that the findings in 2+1 dimensions are a good guide to what would happen in the real world. We will now find out what happens to the Bekenstein-Hawking temperature of scalar and
vector particles from a BBH.

This paper is organized as follows. In Sec. 2, we introduce the geometric
and thermodynamic features of BBH spacetime. In Sec. 3, we study
the Proca equation to find a massive boson in this geometry. Then, we
employ the HJ method with the separation-of-variables technique to
obtain the BHR level of the BBH. Then, in Sec. 4, we repeat the calculations
for the radiating scalar particles under the effect of quantum
gravity. We compute the corrected Bekenstein-Hawking temperature. Finally, in Sec. 5, we obtain the total number of outgoing Bekenstein-Hawking particles, and the paper ends with our conclusions in Sec. 6.

\section{Backreacted Black Hole}

The BBH is constructed from a BTZ BH surrounded by a conformal scalar field. For this reason, one calculates the stress-energy
tensor $<T_{\mu\nu}>$ by considering the transparent boundary conditions 
at infinity; the approximate solution coincides, on a finite domain, with the exact solution \cite{bbh0,bbh1,bbh2}. Transparent boundary conditions, whose name comes from the fact that they are designed to be transparent to outgoing perturbations, have the attribute of being nonreflecting. Transparent boundary conditions were originally designed for the Schrodinger equations but have also been applied to the wave equation \cite{trans} and the conformal scalar field \cite{bbh0,bbh1,bbh2}.
The semiclassical equations are used to find the O($\hbar$) correction to the BBH geometry. 
\begin{equation}
G_{\mu\nu}+\Lambda g_{\mu\nu}=\kappa<T_{\mu\nu}>.
\end{equation}
The exact solution of the metric of the BBH in the presence of a conformally
coupled scalar field in three dimensions can be expressed as~\cite{bbh0}:
\begin{equation}
ds^{2}=-f(r)dt^{2}+f^{-1}(r)dr^{2}+r^{2}d\varphi^{2},\label{eq:metric}
\end{equation}

where 
\begin{equation}
f(r)=\left(\frac{r}{l^{2}}^{2}-M-\frac{2l_{p}F(M)}{r}\right).\label{metricf}
\end{equation}
It is noted that $l^{-2}$ is the cosmological constant, the M is the mass of the BBH and $l_{p}$
is the length of the Planck.
Note that for transparent boundary conditions $F(M)$ is defined as follows \cite{bbh0}: 
\[
F(M)=
\]
\begin{equation}
\frac{M^{3/2}}{2\sqrt{2}}\sum_{n=1}^{\infty}e^{-in\delta}\frac{Cosh[2\pi n\sqrt{M}]+3}{(Cosh[2\pi n\sqrt{M}]-1)^{3/2}},
\end{equation}
where $\delta$ is an arbitrary phase. For a field with periodic boundary conditions $\delta=0$ (bosons), while for fields with antiperiodic boundary conditions $\delta=\pi$ (fermions).

The BTZ BH is recovered at $F(M)=0$ for $M\gg1$. Furthermore, the
metric has an event horizon which is located at 
\begin{equation}
r_{h}=\sqrt{M}l+\frac{l_{p}F(M)}{M}
\end{equation}

\begin{equation}
T_{H}=\frac{f^{\prime}(r_{h})}{4\pi}=\left[\frac{\sqrt{M}}{2\pi l}+\frac{l_{p}F(M)}{\pi Ml^{2}}\right]\label{10n}
\end{equation}

Since the perturbative expansion has no corrections from graviton loops, only radiative corrections are obtained from quantum excitations of the matter fields, because the quantum gravity in 2+1 dimensions is renormalizable and finite. Furthermore, it is observed that the temperature is linear in $F(M)$. The corrections are removed when the $F(M ) \approx e^{- \pi M}$ for a large M, so  the back reaction is more dominant for small masses than for the Planck mass. The other important point is that $<T_{\mu\nu}>$, which depends on the ratio $l_{p}/l$, causes corrections so that for a small cosmological constant, we can ignore the perturbation of the geometry produced by radiation fields.

\section{Bekenstein-Hawking Radiation of Vector Particles From BBH}

To calculate the BHR of the tunneling vector particles from the
BBH, the Proca equation is used on the BBH geometry. The massive vector
particles is described by the Proca equation with the wave function
$\psi$ is given by \cite{aa1,aa2,aa3,aa4,aa5,aa6},

\begin{equation}
\frac{1}{\sqrt{-g}}\partial_{\mu}(\sqrt{-g}\psi^{\nu\mu})+\frac{m^{2}}{\hbar^{2}}\psi^{\nu}=0,\label{14n-1}
\end{equation}
in which

\begin{equation}
\psi_{\nu\mu}=\partial_{\nu}\psi_{\mu}-\partial_{\mu}\psi_{\nu}.\label{15n}
\end{equation}
To leading order in the energy we can neglect the effects of the self-gravitation of the particle \cite{krs1}.
Then we solve the Proca equation on the background of the BBH,
\[
\frac{f}{r}\left\{\partial_r\left[\sqrt{\frac{fr^2}{f}}\left(\partial_t\psi_1-\partial_r\psi_0\right)
\right]+\partial_\varphi\left[\frac{1}{fr}\left(\partial_t\psi_2-\partial_\varphi\psi_0\right)\right]\right\}
+ \frac{m^2}{\hbar^2}\psi_0=0,
\]
\begin{equation}
\frac{1}{{fr}}\left\{\partial_t\left[\sqrt{\frac{f^2r^2}{f}}\left(\partial_t\psi_1-\partial_r\psi_0\right)
\right]+\partial_\varphi\left[\frac{f}{r}\left(\partial_r\psi_2-\partial_\varphi\psi_1\right)\right]\right\}
+ \frac{m^2}{\hbar^2}\psi_1=0,
\label{5}
\end{equation}
\[
r\left\{\partial_t\left[\frac{1}{{fr}}\left(\partial_t\psi_2-\partial_\varphi\psi_0\right)
\right]+\partial_r\left[\sqrt{\frac{f^2}{r^2}}\left(\partial_\varphi\psi_1-\partial_r\psi_2\right)\right]\right\}
+ \frac{m^2}{\hbar^2}\psi_2=0.
\] \label{sol}

It is assumed that the solution exists in the form of the vector functions:
\begin{equation}
\psi_{\nu}=c_{\nu}\exp\left[\frac{i}{\hbar}S(t,r,\varphi)\right],\label{16n-1}
\end{equation}
where $c_{\nu}={c_{0},c_{1},c_{2}}$  are some arbitrary constants,
with the help of action
\begin{equation}
S(t,r,\varphi)=S_{0}(t,r,\varphi)+\hbar S_{1}(t,r,\varphi)+\hbar^{2}S_{2}(t,r,\varphi)+... .\label{17}
\end{equation}

Then the above equations become 

\[
f\left[c_0(\partial_rS_0)^2-c_1(\partial_rS_0)(\partial_tS_0)\right]+
\frac{1}{r^2}\left[c_0(\partial_\varphi S_0)^2-c_2(\partial_\varphi S_0)(\partial_tS_0)\right]+m^2c_0=0,
\]
\begin{equation}
\frac{1}{f}\left[c_0(\partial_rS_0)(\partial_tS_0)-c_1(\partial_tS_0)^2\right]+
\frac{1}{r^2}\left[c_1(\partial_\varphi S_0)^2-c_2(\partial_\varphi S_0)(\partial_r S_0)\right]+m^2c_1=0,
\label{8}
\end{equation}
\[
\frac{1}{f}\left[c_0(\partial_\varphi S_0)(\partial_tS_0)-c_2(\partial_tS_0)^2\right]+
f\left[c_2(\partial_r S_0)^2-c_1(\partial_\varphi S_0)(\partial_r S_0)\right]+m^2c_2=0.
\]
Using the WKB approximation,  the action can be chosen  at the leading order in $\hbar$ as 
\begin{equation}
S_{0}(t,r,\theta)=-Et+W(r)+J\varphi+\Bbbk.\label{18}
\end{equation}
Here the energy is defined by $E$ and the angular momentum of the
spin-1 vector particles is defined by $J$, furthermore $\Bbbk$ is
a constant. Then we use the Eq.s (\ref{18}) inside the solutions of the Eq.(\ref{8}) and considering the leading order in $\hbar$. Then $3\times3$
matrix (let us say $\smiley$ matrix) is obtained: \ $\smiley\left(c_{0},c_{1},c_{2}\right)^{T}=0$.
It is noted that the superscript $T$ means the transition to the
transposed vector. So, the non-zero components of the matrix of $\smiley$
are calculated as follows

\begin{align}
\smiley_{11} & =f(\partial_rW)^2+\frac{(\partial_\varphi J)^2}{r^2}+m^2,\nonumber \\
\smiley_{12} & =fE\partial_rW,\nonumber \\
\smiley_{13} & =\frac{E\partial_\varphi J}{r^2},\ \nonumber \\
\smiley_{21} & = \frac{E\partial_rW}{f}, \nonumber \\
\smiley_{22} & =\frac{E^2}{f}-\frac{(\partial_\varphi J)^2}{r^2}-m^2 \nonumber \\
\smiley_{23} & =\frac{(\partial_rW)\partial_\varphi J}{r^2}\nonumber \\
\smiley_{31} & = \frac{E\partial_\varphi J}{f} \nonumber \\
\smiley_{32} & = f(\partial_r W)\partial_\varphi J \nonumber \\
\smiley_{33} & =\frac{E^2}{f}-f(\partial_r W)^2-m^2 \label{19}
\end{align}

The condition of the finding nontrivial solutions of any linear equations
( $det\smiley=0$ ) gives

\begin{equation}
m^2\left[f(\partial_rW)^2-\left(\frac{E^2}{f}-\frac{(\partial_\varphi J)^2}{r^2}-m^2\right)\right]^2=0.\label{22}
\end{equation}

Then the solution of the above equation for the radial function $W(r)$ yields

\begin{equation}
W_{\pm}(r)=\pm \int\sqrt{\frac{E^2-f\left(m^2+\frac{(\partial_\varphi J)^2}{r^2}\right)}{f^2}}dr.\label{23-1}
\end{equation}
It is noted that the $W_{+}(r)$ and $W_{-}(r)$ show that the vector
particles move away from the BBH and move towards to the BBH. Moreover,
there are poles located at horizon $ r=r_{h}$ the imaginary part of $W_{\pm}(r_{h})$
can be calculated by using the complex path integration method. \cite{aa6,aa7,aa8}
Then, the integral becomes

\begin{equation}
ImW_{\pm}(r_{h})=\pm\frac{\pi}{f^{\prime}(r_{h})}E.\label{24n}
\end{equation}

Then we obtain the probabilities of the vector particles tunnels through the horizon out/in:

\begin{align}
P_{emission} & =e^{-\frac{2}{\hbar}ImS_{+}}=e^{\left[-\frac{2}{\hbar}(ImW_{+}+Im\Bbbk)\right]},\label{26n}\\
P_{absorption} & =e^{-\frac{2}{\hbar}ImS_{-}}=e^{\left[-\frac{2}{\hbar}(ImW_{-}+Im\Bbbk)\right]}.\label{27}
\end{align}

The ingoing vector particles must have the $P_{absorption}=1$ which
mean that their chance to fall inside is 100\% in agreement with
the definition of the BH \cite{aa5,aa6,aa7}. Consequently, we can
choose the $Im\Bbbk=-ImW_{-}$ , then it becomes $W_{+}=-W_{-}$ and
we can calculate the tunneling rate of the vector particles as

\begin{equation}
\Gamma=P_{emission}=\exp\left(-\frac{4}{\hbar}ImW_{+}\right)=\exp\left(-\frac{4\pi}{f^{\prime}(r_{h})}E\right).\label{28}
\end{equation}
Note that the Bekenstein-Hawking temperature of the BBH is recovered by using
relation between the tunneling rate and the Boltzman factor $e^{-\frac{E}{T}}$.
Hence, the Bekenstein-Hawking temperature of BBH is \cite{bbh0,bbh1,bbh2} 
\begin{equation}
T_{H}=\left[\frac{\sqrt{M}}{2\pi l}+\frac{l_{p}F(M)}{\pi Ml^{2}}\right].\label{29}
\end{equation}
Hence, we obtain the correct Bekenstein-Hawking temperature of the BBH using the tunneling method for the spin-1 particles.

\section{Bekenstein-Hawking Radiation of Scalar Particles From BBH with the Effect of the Quantum Gravity}

In this section, we check the effects of quantum gravity on the BHR of scalar particles from a BBH using the generalized uncertainty principle (GUP). One may ask how we can extend the quantum mechanics, considering the gravitational interactions. The answer is a quantum theory of gravity, which is the biggest problem in theoretical physics. One common feature among various quantum gravity theories, such as
string theory, loop quantum gravity, and noncommutative geometry,
is the existence of a minimum measurable length \cite{gp1,gp2,gp3,gp4,gp5,gp6,gp7,gp8,gp9}.
An effective model for realizing the minimum length is the GUP, based on which the first generalized
uncertainty relationship was proposed by \cite{gp3}, to solve the problem of quantum gravity.

First, we solve the Klein-Gordon (KG) equation under the effect of the GUP on the background of the BBH to find the BHR. The commutation relationship is modified by
using quantum gravity as follows \cite{hu3,gp2,gp3,gp4}:

\begin{equation}
\lbrack x_{i},p_{j}]=i\hbar(1+\alpha^{2}p^{2})\delta_{ij},
\end{equation}
and the GUP is derived as follows 
\begin{equation}
\Delta x\Delta p\geq\frac{\hbar}{2}\left\{ 1+\alpha^{2}(\Delta p)^{2}\right\} .
\end{equation}
It is noted that $\alpha=\alpha_{0}/(m_{p}^{2})=l_{p}^{2}/\hbar^{2}$
is a small value, $m_{p}=\frac{\hbar}{l_{p}}$ is the Planck mass,
$l_{p}$ is the Planck length ($\sim10^{-35}m$) and $\alpha_{0}<10^{34}$
is a dimensionless parameter.

Quantum gravity effects the KG equation which is the relativistic
wave equation for the scalar particles, because the position, momentum
and energy operators are modified due to the GUP respectively as follows

\begin{equation}
x_{i}=x_{oi},
\end{equation}

\begin{equation}
p_{i}=p_{0i}(1+\alpha^{2}p^{2})
\end{equation}

and

\begin{equation}
\varepsilon=E(1+\alpha^{2}E^{2}).
\end{equation}
Furthermore, the frequency is also generalized as

\begin{equation}
\bar{\omega}=E(1-\alpha E^{2}),\label{eq4}
\end{equation}

\noindent with the energy operator $E=i\hbar\partial_{0}$. One can
calculate the square of momentum operators up to order $\alpha^{2}$
as 
\begin{eqnarray}
p^{2} & = & -\hbar^{2}[1-\alpha^{2}\hbar^{2}\partial_{j}\partial^{j}]\partial_{i}[1-\alpha^{2}\hbar^{2}\partial_{j}\partial^{j}]\partial^{i}\nonumber \\
 & = & -\hbar^{2}[\partial_{i}\partial^{i}-2\alpha^{2}\hbar^{2}(\partial_{j}\partial^{j})(\partial_{k}\partial^{k})]+(\alpha^{4}),\label{sq}
\end{eqnarray}

\noindent where in the last step, we only keep the leading order term
of $\alpha$.

\noindent Therefore, the generalized KG equation with the wave function
$\Psi$ can be written as \cite{hu3}

\begin{equation}
-(i\hslash)^{2}\partial^{t}\partial_{t}\Psi=\left[(i\hslash)^{2}\partial^{i}\partial_{i}+m_{p}^{2}\right]\left[1-2\alpha\left((i\hslash)^{2}\partial^{i}\partial_{i}+m_{p}^{2}\right)\right]\Psi.\label{14n}
\end{equation}
Herein, we substitute the ansatz for the semiclassical wave function
$\Psi$ of the scalar particles 
\begin{equation}
\Psi=Ce^{\frac{i}{\hbar}S(t,r,\theta)}\label{16n}
\end{equation}
where C is the constant, into the generalized KG equation (Eq.\ref{14n})
with the BBH \ metric Eq.(\ref{eq:metric} ) which is the background
of scalar particle motion. Then, the differential equation for the
action S is calculated as follows

\begin{eqnarray}
 &  & \frac{1}{f}\left[\left(d_{t}S\right)^{2}=f(d_{r}S)^{2}+\frac{1}{r^{2}}(d_{\theta}S)^{2}+m_{p}^{2}\right]\label{kg}\\
 &  & \times\left[1-2\alpha\left(f(d_{r}S)^{2}+\frac{1}{r^{2}}(d_{\theta}S)^{2}+m_{p}^{2}\right)\right]\nonumber 
\end{eqnarray}
The separation of variables are used to solve the generalized KG equation
after the Eq.(\ref{kg}) is expanded into the lowest order of $\hbar$
\begin{equation}
S(t,r,\theta,\varphi)=-Et+W(r)+j(\theta)+\varrho,\label{alice}
\end{equation}
where $\varrho$is the constant. Then we substitute Eqn. (\ref{alice})
into Eq. (\ref{kg}) to solve for the W(r). Then the radial part of
the scalar wave function is found that

\begin{equation}
W\pm=\int dr\frac{1}{f}\frac{\sqrt{E^{2}-m_{p}^{2}(1-2\alpha m_{p}^{2})f}}{\sqrt{1-2\alpha m_{p}^{2})}},
\end{equation}
where the positive and negative $\pm$\ signatures are for the outgoing
and ingoing scalar particles. To solve this integral, after using
the residue method around the pole at the horizon we obtain the solution
\begin{equation}
W\pm=i\pi\frac{E}{f^{\prime}(r_{h})\sqrt{1-2\alpha m_{p}^{2})}}.
\end{equation}
Herein, similarly to the previous section, we use the fact that the
probability of ingoing particles to 100\% $(P_{absorption}=1)$. Thus,
the tunneling rate is calculated for the scalar particles with the
effect of the quantum gravity as

\begin{equation}
\Gamma=P_{emission/absorption}=e^{-\frac{4}{\hbar}ImW_{+}}=e^{-\frac{4\pi i}{f^{\prime}(r_{h})\sqrt{1-2\alpha m_{p}^{2})}}E}.
\end{equation}
Now, it is easy to recover Bekenstein-Hawking temperature for the scalar particles
with the effect of quantum gravity

\begin{equation}
T_{H}=\frac{f^{\prime}(r_{h})\sqrt{1-2\alpha m_{p}^{2}}}{4\pi}
\end{equation}
\[
T_{H}=\left[\frac{\sqrt{M}}{2\pi l}+\frac{l_{p}F(M)}{\pi Ml^{2}}\right]\sqrt{1-2\alpha m_{p}^{2}}.
\]

It is easilty observed that when we choose $\alpha=0,$ it is equal
to the original result of Bekenstein-Hawking temperature. Hence, the BHR of the BBH with the effect of the quantum gravity has remnants.

\section{Total Number Of Tunneling Massless Bekenstein-Hawking particles}

In this section, we calculate the estimation of the total number of
massless quanta emitted by the BBH. One shows that the total number
of quanta emitted by the BBH is proportional to the square of the BBH's
initial mass in Planck units. Firstly we introduce the Planck's law
of black-body radiation for two space dimensions to calculate the
spectral luminosity density of an ideal black body as follows ($8\pi^{2}=\hbar=G=k_{B}=c=1$.)
\cite{muck,viser,val} 
\begin{equation}
dL=\frac{\omega^{2}}{\exp{\omega/T}-1}d\omega dA.\label{lum.spec.density}
\end{equation}
Note that $\omega$, $A$ and T are the energy, surface area and the
temperature, respectively. The result of the integration of the Eq.\eqref{lum.spec.density}
is the Stefan-Boltzmann law which is stated that the power emitted
per unit area of the surface of a black hole is directly proportional
to the 4th power of its temperature \cite{muck,viser}. After we take
integral of the Eq.\eqref{lum.spec.density}, for two space dimensions,
the luminosity is found as 
\begin{equation}
L\eqsim AT^{3}.\label{stefan.boltzmann}
\end{equation}
Then the emission rate of the emitted quanta is obtained as 
\begin{equation}
\Gamma=\frac{1}{\omega}\frac{dL}{d\omega}=\int\frac{gA\omega^{2}d\omega}{\exp{\omega/T}-1}~,\label{num.flux.spec.density}
\end{equation}
where g the number of radiating degrees of freedom,
\begin{equation}
\Gamma\eqsim AT^{2}.\label{num.flux}
\end{equation}
Now, we recall the Bekenstein-Hawking temperature of the BBH and the area of
the BBH in 2+1-dimension as 
\begin{equation}
T\eqsim\frac{\sqrt{M}}{l}+\frac{l_{p}F(M)}{Ml^{2}},\label{schwarzschild}
\end{equation}
and
\[
A\eqsim r_{h}\eqsim\left[\sqrt{M}l+\frac{l_{p}F(M)}{M}\right].
\]
Once shows that the mass loss rate is related with the luminosity
as follows 
\begin{equation}
\frac{dM}{dt}=-L=-\frac{M^{2}}{l^{2}}.\label{dM.dt}
\end{equation}
Then the evaporation time of the BTZ BH is obtained as

\begin{equation}
t_{evaporation}=-\frac{l^{2}}{M}\label{tev}
\end{equation}
After that we calculate the emission rate of the Bekenstein-Hawking particles
\begin{equation}
\Gamma=\frac{M^{3/2}}{l}~.\label{BH.num.flux}
\end{equation}
The total number of the outgoing Bekenstein-Hawking particles are obtained by
following relation \cite{muck,viser} 
\begin{equation}
\frac{dN}{dM}=\frac{\Gamma}{\frac{dM}{dt}}=-\frac{l}{\sqrt{M}},\label{dN.dM}
\end{equation}
and it is found as 
\begin{equation}
N=-2l\sqrt{M}.\label{N.BH}
\end{equation}
Note that it does not depend on the spin of the particles so both
vector and scalar particles radiating from BBH with the same number
of particles \cite{muck,viser}.

For the case of scalar particles with the effect of quantum gravity,
the total number of the outgoing Bekenstein-Hawking particles is 
\begin{equation}
N=\frac{-2l\sqrt{M}}{\sqrt{(1-2\alpha m_{p}^{2})}}.\label{N.BH-1}
\end{equation}

The total number of Bekenstein-Hawking particles ($N$) increases with the effect
of the quantum gravity constant $\alpha$. However, at some point, $N$ becomes zero, and no particles are emitted. The results of the calculation of the total number of tunneling massless Bekenstein-Hawking particles are compatible with recent studies on the information paradox \cite{muck}. Also, one can calculate the number of gravitons in the BH quantum $N$ portrait with similar conclusions\cite{viser2,D1,D2}. In this paper, we follow the same proposal as \cite{muck}, but we investigate it differently for 2+1 dimensions, imagining a BH to be a Bose-Einstein condensate with very large and massive gravitons. To solve the information paradox, BHR is thought of as resulting from a decrease in the condensate with nonthermal properties of order $1/N$ when a two-body interaction occurs and also gives evidence of a quantum $N$ portrait of a semiclassical BH \cite{D1,D2}. One can interpret each Bekenstein-Hawking radiated particle as an information storage unit; in fact, there is an expected link between the entropy and the particle number, so that the BHR can be thought of as sparse in a semiclassical (corpuscular) regime.

\section{Conclusions}
In this paper, first by using the generalized Klein-Gordon
and Proca equations, we investigate the scalar/vector particles tunneling from a BBH and recover the corresponding Bekenstein-Hawking temperatures. For this purpose, first we use the Proca equation for the tunneling spin-1 (vector) particles on the background of BBH spacetime. Using the WKB approximation to the Proca equation, we find the set of field equations. Then we use the Hamilton-Jacobi ansatz to solve these equations. To solve these equation, we take the determinant of coefficient matrix as zero and we expand the functions $f(r)$ in Taylor's series near horizon to find the radial wave equation $W(r)$ using the complex path integral. Then we use the surface gravity and calculate the probability of tunneling of spin-1 particles from the BBH. Using the Boltzmann formula, we derive the corresponding Bekenstein-Hawking temperature. It is worth while to mention here that the self- gravitating effects have been neglected, however there is a back-reaction effects because of the BBH geometry. Moreover, we calculate the Bekenstein-Hawking temperature only in a leading order term. Hence, we can conclude that Bekenstein-Hawking temperature is not dependent of the types of particles and also the tunneling probabilities are same. Therefore, their corresponding Bekenstein-Hawking temperatures are same for all kinds of particles. 

Second, we use the effect of GUP that the existence of a minimal length leads to the modification of the Heisenberg uncertainty principle on the tunneling scalar particles. The GUP contains an additional quadratic term in momentum in addition to a minimal length. After we generalize the  Klein-Gordon equation using the effect of the GUP, we focus on the Hamilton-Jacobi method to determine the tunneling probability of the scalar particles. Again we use the WKB approximation and Hamilton-Jacobi ansatz in the tunneling formalism and calculate the imaginary part of the action in order to obtain the Bekenstein-Hawking temperature. Hence, it is shown that if the GUP is used, then the Bekenstein-Hawking temperature of the tunneling scalar particle at the event horizon differ from the original case and the Bekenstein-Hawking temperature has a nonthermal feature. The backreactions on the black hole have also similar nonthermal effects on the Bekenstein-Hawking temperature. It is concluded that using the GUP, decreases the backreacted effects on the Bekenstein-Hawking temperature. However, the GUP effects are not sufficient to extinguish the Bekenstein-Hawking temperature.

Scalar and vector particles radiate
from a BBH with equivalent energies if the GUP is not used. After one uses the generalised Klein-Gordon equation with the effect of GUP, the corrected temperature decreases with the effects of quantum gravity, and at some point, remnants are left. Then, we calculate the total number of emitted Bekenstein-Hawking
particles from the special case of a BBH, which is a BTZ BH. Also, we check
the effects of quantum gravity on the total number of emitted particles
from the BH. It is shown that the emitted Bekenstein-Hawking particles are information-carrying units. This indicates a corpuscular
interpretation instead of an undulatory one, and when the BH collapses, its unitary property is preserved, and the BH evaporates. Hence, the effect
of quantum gravity balances the classical tendency of rising temperature, 
and there exist remnants. In summary, it is very rare for the BHR to be extremely diluted. The BHR particles are discrete during propagation and also appear as particles later on so that the BHR particles appears to be a particle feature.

\begin{acknowledgments}
This work was supported by the Chilean FONDECYT Grant No. 3170035.
\end{acknowledgments}

\section*{Competing Interests}
 The author declare that there is no conflict of interest regarding the publication of this paper.

\end{document}